\documentclass[a4paper,11pt]{article}
\usepackage{pos}
\usepackage[capitalize]{cleveref}
\usepackage{amsmath}
\usepackage{bm} 
\usepackage{subcaption}
\usepackage[section]{placeins}

\title{Comparison of lattice QCD results for inclusive semi-leptonic decays of \texorpdfstring{$B$}{} mesons with the OPE}
\ShortTitle{OPE and LQCD}

\author[a]{Paolo Gambino}
\author[b]{Shoji Hashimoto}
\author*[a,c]{Sandro M\"achler}

\affiliation[a]{Dipartimento di Fisica, Universit`a di Torino \& INFN, Sezione di Torino,\\
  Via Pietro Giuria, I-10125 Torino, Italy}

\affiliation[b]{Theory Center, Institute of Particle and Nuclear Studies,\\
High Energy Accelerator Research Organization (KEK), Tsukuba 305-0801, Japan\\
and School of High Energy Accelerator Science,\\
The Graduate University for Advanced Studies (SOKENDAI),\\
Tsukuba 305-0801, Japan}

\affiliation[c]{Physikinstitut, Universit\"at Z\"urich, Winterthurerstrasse 190, CH-8057 Z\"urich, Switzerland}

\emailAdd{paolo.gambino@unito.it}
\emailAdd{shoji.hashimoto@kek.jp}
\emailAdd{sandro.machler@unito.it}

\abstract{A recently proposed approach to compute the inclusive semi-leptonic decay rate on the lattice allows for the calculation of various quantities (differential distributions and moments) in different subchannels. We systematically compare the lattice QCD results for an unphysically light bottom quark with OPE-based predictions, including leading order perturbative and $\mathcal{O}\left(1/m^3 \right)$ power corrections, and explore possible strategies to decrease the systematic uncertainty.\newline
\newline
Preprint number: KEK-CP-0383}

\FullConference{%
 The 38th International Symposium on Lattice Field Theory, LATTICE2021
  26th-30th July, 2021
  Zoom/Gather@Massachusetts Institute of Technology
}

\begin{document}
\maketitle

\section{Introduction}
  The study of decays of $b$ quarks is an immensely fascinating and promising area of research. Semi-leptonic decays of $b$ quarks are particularly interesting because they can be used to determine the CKM matrix element $V_{cb}$. In fact there is an unresolved tension between the values of $V_{cb}$ determined from inclusive and exclusive decays of $B$ mesons \cite{Gambino:2019sif,ParticleDataGroup:2020ssz,HFLAV:2019otj}.

  The term inclusive here refers to the fact that the final hadronic state is not fully known. Instead the only available information about its particle content is that it contains a charm quark. The method of choice for describing these decays is the use of the optical theorem followed by performing an operator product expansion (OPE) \cite{Chay:1990da, Manohar:1993qn,Blok:1993va}. In this approach observables are the total rate and smeared differential distributions, given in terms of a double expansion in $\Lambda_{\mathrm{QCD}}/m_b $ and $\alpha_s$. A generic observable $M_i$ can then be written as
  \begin{align}\label{eq:InclusiveObservable}
      M_i &= M_i^{(0,0)} + \frac{\alpha_s}{\pi} M_i^{(1)} + \frac{\mu_\pi^2}{m_b^2} \left( M_i^{(\pi,0)} + \frac{\alpha_s}{\pi} M_i^{(\pi,1)} \right) \nonumber \\
      &\phantom{=} + \frac{\mu_G^2}{m_b^2}\left(M_i^{(G,0)} + \frac{\alpha_s}{\pi} M_i^{(G,1)} \right) + \frac{\rho_D^3}{m_b^3}M_i^{(D,0)} + \frac{\rho_{LS}^3}{m_b^3} M_i^{(LS,0)} + \cdots.
  \end{align}
  At any order in this expansion one encounters a number of non-perturbative matrix elements. In \cref{eq:InclusiveObservable} they are denoted by $\mu_\pi^2, \mu_G^2, \rho_D^3 $ and $ \rho_{LS}^3 $, where for instance
  \begin{equation}
      \mu_\pi^2 = \frac{1}{M_B}\left\langle B \left\vert \overline{b}_v \left(i \stackrel{\rightarrow}{D} \right)^2 b_v \right\vert B \right\rangle, \qquad \mu_G^2 = \frac{1}{M_B} \left\langle B \left\vert \overline{b}_v \frac{i}{2} \sigma_{\mu\nu}G^{\mu\nu} b_v \right\vert B \right\rangle.
    \end{equation}
  These matrix elements cannot be computed from first principles, instead they are treated as free parameters and extracted from experimental data.

  Recently a novel method for computing inclusive observables in lattice QCD has been proposed \cite{Hashimoto:2018gld}. In particular in the light of the aforementioned tension between the inclusive and exclusive determinations of $V_{cb}$ and the flavor anomalies \cite{LHCb:2021trn}, such an independent computation of inclusive observables can serve as an important check of the OPE approach. To this end a first numerical study has been performed in \cite{Gambino:2020crt}. We build on this study and expand it, including further terms in the heavy quark expansion and examining additional observables.

  This paper is organized as follows. In \cref{sec:InclusiveDecays} we introduce the OPE framework used to study inclusive decays of $B$ mesons, followed by a discussion of the observables considered in \cref{sec:Observables} and in \cref{sec:Comparison} the OPE results are confronted with lattice data obtained by the JLQCD collaboration. 

\section{Inclusive decays of \texorpdfstring{$B$}{} mesons}
  \label{sec:InclusiveDecays}

  We consider the decay
  \begin{equation}
    B\left(m_B,\bm{0} \right) \rightarrow X_c\left(\omega, -\bm{q} \right)\ell\left(E_\ell, \bm{p}_\ell \right)\overline{\nu}_\ell \left(q_0 -E_\ell, \bm{q}-\bm{p}_\ell \right)
  \end{equation}
  in the rest frame of the $B$ meson. In the triple differential width for this decay the hadronic and leptonic parts factorize into a leptonic tensor $L_{\mu\nu}$ and a hadronic tensor $W^{\mu\nu} $ as \cite{Manohar:1993qn,Blok:1993va}
  \begin{equation}
    \frac{\mathrm{d}^3\Gamma}{\mathrm{d}\bm{q}^2\mathrm{d}\omega \mathrm{d}E_\ell} = \frac{G_\mathrm{F}^2 \left\vert V_{cb} \right\vert^2}{8\pi^3} L_{\mu\nu} W^{\mu\nu}.
  \end{equation}
  For massless final state leptons the leptonic tensor is given by
  \begin{equation}
    L^{\mu\rho} = p_\ell^\mu p_{\overline{\nu}}^\rho - p_\ell \cdot p_{\overline{\nu}} g^{\mu\rho} + p_\ell^\rho p_{\overline{\nu}}^\mu +i \epsilon^{\mu\alpha\rho\beta}p_{\ell,\alpha} p_{\overline{\nu},\beta}
  \end{equation}
  while the hadronic tensor is defined as
  \begin{align}
    W^{\mu\rho} \left(p,q \right) = \sum_{X_c} \left(2\pi \right)^3 \delta^{(4)}\left(p-q-r \right) \frac{1}{2E_{B_s}} \left\langle \overline{B}_s(\bm{p})\left\vert J^{\mu \dagger} \right\vert X_c \left(\bm{r} \right) \right\rangle \left\langle X_c\left(\bm{r} \right) \left\vert J^{\rho \phantom{\dagger}} \right\vert \overline{B}_s \left(\bm{p} \right) \right\rangle.
  \end{align}
  Here the sum runs over all possible final states $X_c$ containing a charm quark and the electroweak current relevant for this decay mode is given by $J^\mu = \left(V - A \right)^\mu = \overline{c}\gamma^\mu \left(1-\gamma_5 \right) b $. In the following we will keep the contributions from $VV$ and $AA$ products of currents separate and distinguish between polarizations parallel and perpendicular to the $W$ boson's three-momentum $\bm{q}$. There is of course also a mixed product of an axialvector and a vector current which will be included in a future study.

  The hadronic tensor can be decomposed into Lorentz invariant structure functions as
  \begin{align}
    W^{\mu\nu} = &-g^{\mu\nu} W_1\left(q_0, q^2 \right) + v^\mu v^\nu W_2\left(q_0, q^2 \right) -i \epsilon^{\mu\nu\alpha\beta}v_\alpha \hat q_\beta W_3\left(q_0, q^2 \right),
  \end{align}
  where we have used the fact that we consider massless leptons to eliminate two additional Lorentz structures. Through the optical theorem the functions $W_i$ are related to the structure functions of the forward scattering matrix element
  \begin{equation}
    4iT = \frac{1}{2 M_B} \int \mathrm{d}^4x e^{-iqx} \left\langle \overline{B} \left\vert T\left[J^\dagger\left(x \right)J\left(0 \right) \right] \right\vert \overline{B} \right\rangle
  \end{equation}
  which can be expanded into a series of local operators, resulting in an expansion in powers of $\Lambda_\mathrm{QCD}/m_b $, with corrections starting at $\mathcal{O}\left(\left(\Lambda_\mathrm{QCD}/m_b\right)^2 \right) $ \cite{Chay:1990da}. The structure functions $W_i$ are computed in the OPE up to $\mathcal{O}\left(\left(\Lambda_\mathrm{QCD}/m_b\right)^3 \right) $ and can be found for example in \cite{Colangelo:2020vhu}.

\section{Observables}
  \label{sec:Observables}
  In order to compare the results of the OPE computation to lattice data we study the differential $\bm{q}^2 $ distribution and moments of the final state charged lepton energy at fixed values of $\bm{q}^2 $.

  The total decay width $\Gamma$ is found by integrating the triple differential width over phase space
  \begin{equation}
    \Gamma = \frac{G_\mathrm{F} \left\vert V_{cb} \right\vert^2}{24 \pi^3} \int_{0}^{\frac{\left(m_{B_s}^2-m_{D_s}^2\right)^2}{4m_{B_s}^2}}\mathrm{d} \bm{q}^2 \sqrt{\bm{q}^2} \overline{X}\left(\bm{q}^2 \right) ,
  \end{equation}
  where 
  \begin{equation}\label{eq:OmegaIntegral}
    \overline{X}\left(\bm{q}^2 \right) = 3\int_{\sqrt{m_{D_s}^2+\bm{q}^2}}^{m_{B_s}-\sqrt{\bm{q}^2}} d\omega \int_{\frac{q_0^2-\sqrt{\bm{q}^2}}{2}}^{\frac{q_0+\sqrt{\bm{q}^2}}{2}}\mathrm{d}E_\ell L_{\mu\nu}W^{\mu\nu} \left(E_\ell, \omega,\bm{q}^2 \right)
  \end{equation}
  corresponds to the $\bm{q}^2 $ distribution $\mathrm{d}\Gamma / \mathrm{d}\bm{q}^2 $ up to an overall factor of $\left(G_\mathrm{F} \left\vert V_{cb} \right\vert^2 \left\vert \bm{q} \right\vert \right)/ \left(24 \pi^3 \right) $.

  In order to decrease uncertainties it is useful to consider normalized observables such as the moments of the final state charged lepton energy given by
  \begin{equation}\label{eq:LeptonMomentDefinition}
    \left\langle E_\ell^n \right\rangle \left(\bm{q}^2 \right) = \frac{\int \mathrm{d}\omega \mathrm{d}E_\ell E_\ell^n \frac{\mathrm{d}^3 \Gamma}{\mathrm{d}\bm{q}^2 \mathrm{d}\omega \mathrm{d}E_\ell}}{\int \mathrm{d}\omega \mathrm{d}E_\ell \frac{\mathrm{d}^3 \Gamma}{\mathrm{d}\bm{q}^2 \mathrm{d}\omega \mathrm{d}E_\ell}}.
  \end{equation}
  In the following section we show a comparison between OPE predictions and lattice data for the $\bm{q}^2 $ distribution $\overline{X}$, the first moment of the charged lepton energy and its variance, all at fixed values of $\bm{q}^2$.

\section{Comparison to lattice data}
  \label{sec:Comparison}
  We compare the predictions obtained in the OPE with an expanded version of the set of lattice data that has been used in \cite{Gambino:2020crt}. These data are computed on an ensemble with $2+1$ flavors of M\"obius domain-wall fermions (the ensemble "$\mathrm{M}-ud3-s\mathrm{a} $" in \cite{Nakayama:2016atf}, which has an inverse lattice spacing of $1/a=3.610(9) $ GeV). For the charm and bottom quarks in the valence sector the same lattice formulation is used. The charm quark mass $m_c$ is tuned to its physical value and the $D_s$, $D_s^*$ meson masses are given by $1.98$ GeV and $2.12$ GeV respectively. The bottom quark mass is taken as $2.44 m_c$, which is substantially smaller than the physical $b$ quark mass. The corresponding $B_s$ meson mass is $3.45$ GeV. In this setup the maximum possible spatial momentum in the $B_s \rightarrow D_s \ell \nu_\ell$ decay is $\left(m_{B_s}^2 - m_{D_s}^2 \right)^2/2m_{B_s} \approx 1.16 $ GeV. The lattice volume is $L^3 \times L_t = 48^3\times96$ and the forward-scattering matrix elements have been computed at spatial momenta $\bm{q}$ of $(0,0,0)$, $(0,0,1) $, $(0,1,1) $, $(1,1,1) $ and $(0,0,2) $ in units of $2\pi/L\mathrm{a}$. The number of lattice configurations averaged is 100 and the measurement is performed with four different time-slices. More details on the lattice calculation are presented in \cite{Hashimoto:2017wqo}. 

\begin{figure}[h!]
    \includegraphics[width=\textwidth]{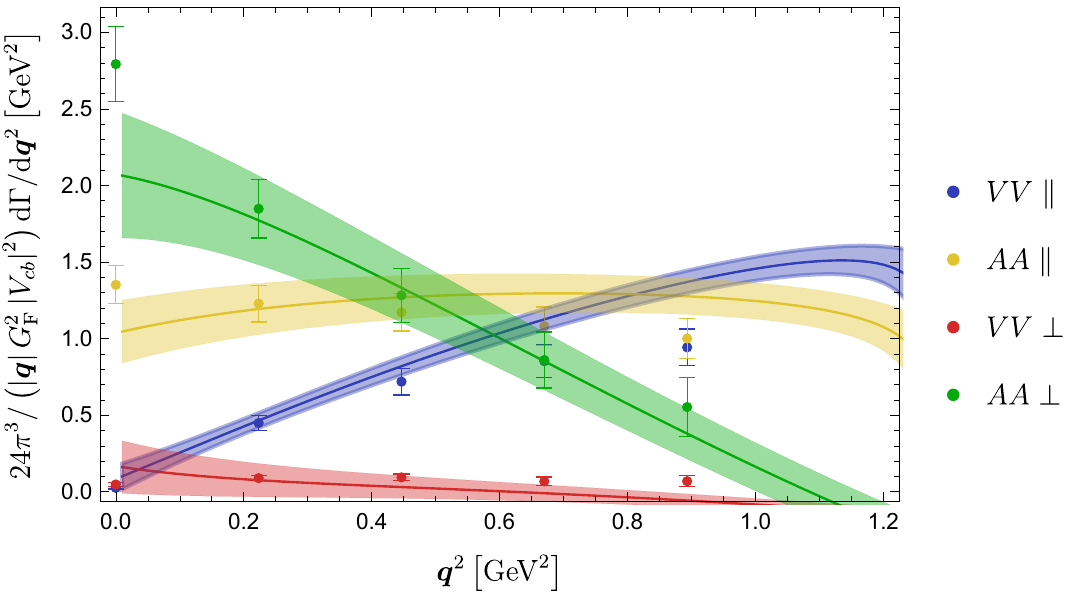}
    \caption{\label{fig:TotalWidthComparison} Contributions of the individual channels to the $\bm{q}^2$ distribution. The dots and solid lines correspond to lattice data and OPE prediction at $\mathcal{O}\left(1/m_b^3,\alpha_s \right) $ respectively. The shaded bands include the experimental uncertainty on the non-perturbative parameters and theoretical uncertainty due to the truncation of the double series.}
  \end{figure}

  The lattice results for the differential $\bm{q}^2 $ distribution are compared with the OPE predictions in \cref{fig:TotalWidthComparison}. As mentioned above the results for the different polarizations parallel and perpendicular to $\bm{q} $ are plotted separately for vector (VV) and axial-vector (AA) contributions.

  The OPE predictions shown in this plot include non-perturbative power corrections up to $\mathcal{O}\left(1/m_b^3 \right) $ and $\mathcal{O}\left(\alpha_s \right) $ perturbative corrections. In the OPE computation we take the $\overline{MS} $ mass $\overline{m}_c (2 \text{ GeV} ) = 1.093 $ GeV for the charm quark and the kinetic mass for the unphysically light bottom quark, $m_b^\mathrm{kin}( 1 \text{ GeV} ) = 2.70(4) $ GeV. The bottom quark mass is tuned to reproduce the $B_s$ meson mass in the lattice simulation using the results of \cite{Gambino:2017vkx}. For the non-perturbative matrix elements we employ the results of the recent semi-leptonic fit in \cite{Bordone:2021oof} even though it refers to a physical $b$ quark and a light spectator. Expressions for the $\mathcal{O}\left(\alpha_s \right) $ corrections can be found in \cite{Aquila:2005hq}. Here they are used with $\alpha_s = 0.32$. An estimate of the theoretical uncertainties and the statistical uncertainties originating in the semi-leptonic fit determining the OPE matrix elements is given in the form of shaded bands around the OPE lines.

  In order to estimate the theoretical uncertainties we assume that the Wilson coefficients of $\mu_\pi^2$ and $\mu_G^2$ are affected by perturbative corrections at the level of $15 \%$, yielding
  $\mu_\pi^2 = 0.477 \pm 0.07155 \text{ GeV}^2 $ and $\mu_G^2 = 0.306 \pm 0.0459 \text{ GeV}^2 $ , while the Wilson coefficients of $\rho_D^3$ and $\rho_{LS}^3$ can effectively be changed by non-perturbative and perturbative corrections at the $25\%$ level, giving $\rho_D^3 = 0.185 \pm 0.04625 \text{ GeV}^3 $ and $\rho_{LS}^3 = -0.13 \pm 0.0325 \text{ GeV}^3  $. Additionally we assign an uncertainty of $6$ MeV to the quark masses and vary $\alpha_s$ by $4\%$. These theoretical uncertainties are then added in quadrature to the statistical ones originating in the semi-leptonic fit.

  In \cref{fig:TotalWidthComparison} we show the combination of theoretical and statistical uncertainties. Because we do not estimate the effect of the unphysically small $b$ quark mass on the OPE matrix elements the bands are slightly underestimated however.

\begin{figure}[h!]
    \includegraphics[width=\textwidth]{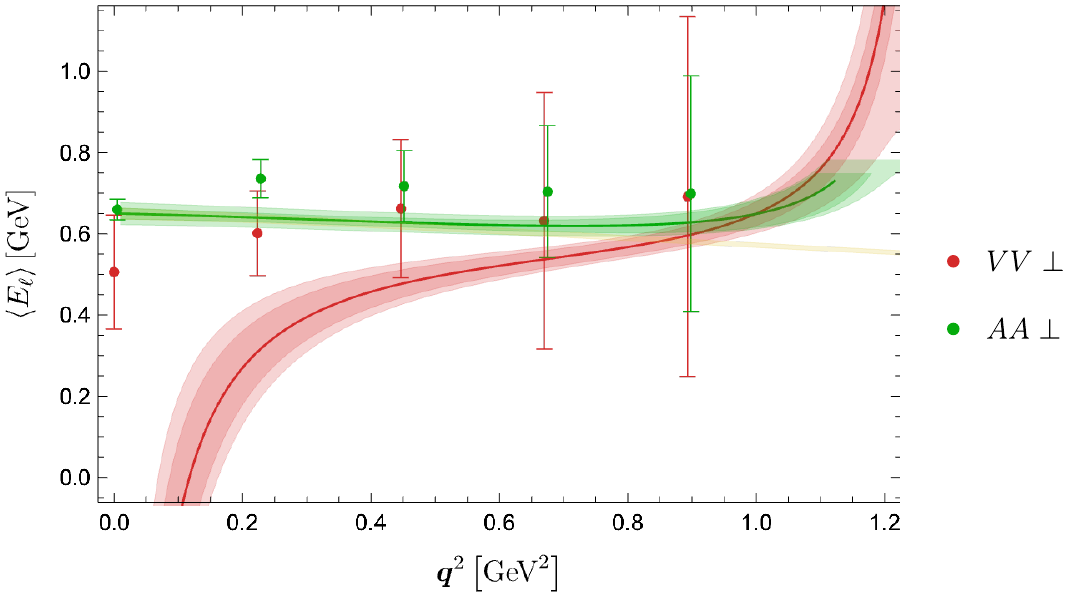}
    \caption{\label{fig:FirstLeptonMomentPlotPerpendicular} First lepton energy moment at fixed $\bm{q}^2$. The dots and solid lines correspond to lattice data and OPE prediction at $\mathcal{O}\left(1/m_b^3,\alpha_s \right) $ respectively. The shaded bands include the experimental uncertainty on the non-perturbative parameters and theoretical uncertainty due to the truncation of the double series.}
  \end{figure}

  Keeping this in mind we find a remarkable agreement between the lattice data and OPE predictions for the differential $\bm{q}^2$ distribution.
  \begin{figure}[h!]
    \includegraphics[width=\textwidth]{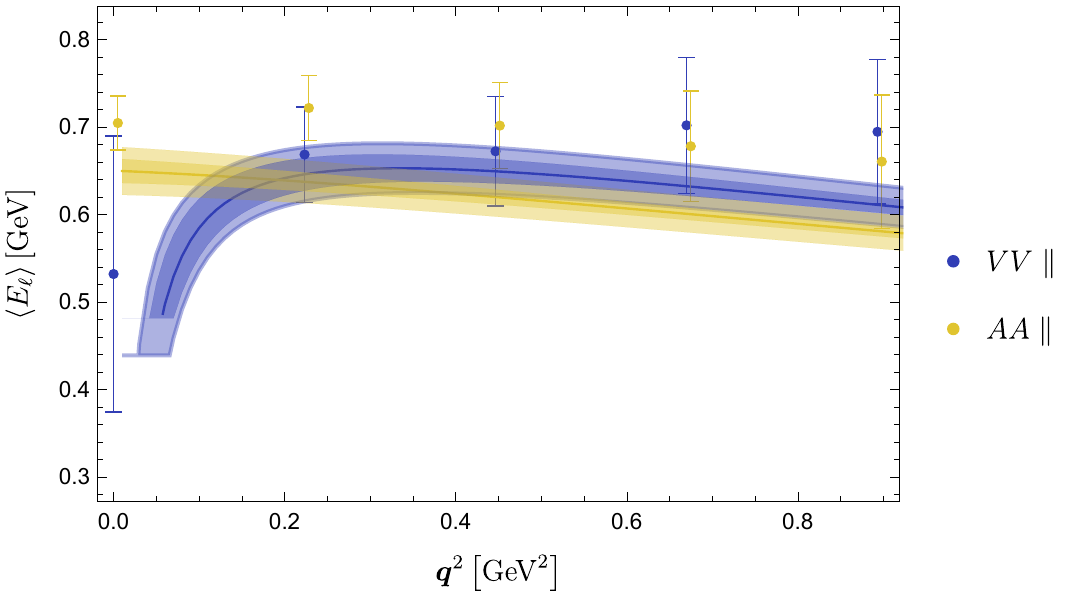}
    \caption{\label{fig:FirstLeptonMomentPlotParallel} First lepton energy moment at fixed $\bm{q}^2$. The dots and solid lines correspond to lattice data and OPE prediction at $\mathcal{O}\left(1/m_b^3,\alpha_s \right) $ respectively. The shaded bands include the experimental uncertainty on the non-perturbative parameters and theoretical uncertainty due to the truncation of the double series.}
  \end{figure}
  The OPE predictions strongly depend on the quark masses. This dependence partially cancels, thus reducing the OPE uncertainties, if one considers normalized quantities such as the moments of the charged lepton energy defined in \cref{eq:LeptonMomentDefinition}. A comparison between lattice data and OPE predictions for the average charged lepton energy at fixed values of $\bm{q}^2$ can be seen in \cref{fig:FirstLeptonMomentPlotParallel,fig:FirstLeptonMomentPlotPerpendicular}. Here we show two kinds of error bands around the OPE predictions. The dark shaded bands correspond to the theory error alone, while the light shaded bands correspond to the combination of theoretical uncertainties and statistical uncertainties as in \cref{fig:TotalWidthComparison}.
  Again we find an excellent agreement between OPE predictions and lattice data. In the $VV \perp$ channel the agreement is worse than in the others. One can make sense of this by noting that for this channel the average charged lepton energy at fixed  $\bm{q}^2$ is given by
  \begin{equation}
    \left\langle E_\ell \right\rangle_{VV \perp} \left(\bm{q}^2 \right) = \frac{\left. \int \mathrm{d}\omega \mathrm{d}E_\ell E_\ell^n \frac{\mathrm{d}^3 \Gamma}{\mathrm{d}\bm{q}^2 \mathrm{d}\omega \mathrm{d}E_\ell}\right\vert_{VV \perp}}{\left. \int \mathrm{d}\omega \mathrm{d}E_\ell \frac{\mathrm{d}^3 \Gamma}{\mathrm{d}\bm{q}^2 \mathrm{d}\omega \mathrm{d}E_\ell}\right\vert_{VV \perp}} = \frac{\left. \int \mathrm{d}\omega \mathrm{d}E_\ell E_\ell^n \frac{\mathrm{d}^3 \Gamma}{\mathrm{d}\bm{q}^2 \mathrm{d}\omega \mathrm{d}E_\ell}\right\vert_{VV \perp}}{\left. \frac{\mathrm{d}\Gamma }{\mathrm{d}\bm{q}^2}\right\vert_{VV \perp}}
  \end{equation}
  and the fact that the denominator $\left. \mathrm{d}\Gamma /\mathrm{d}\bm{q}^2\right\vert_{VV \perp}$ is close to 0, as can be seen in \cref{fig:TotalWidthComparison}. Consequently even a small change in the denominator could have a large effect on $\left\langle E_\ell \right\rangle_{VV \perp} $.
  \begin{figure}[!ht]
    \begin{subfigure}{0.495\textwidth}
      \includegraphics[width=\textwidth]{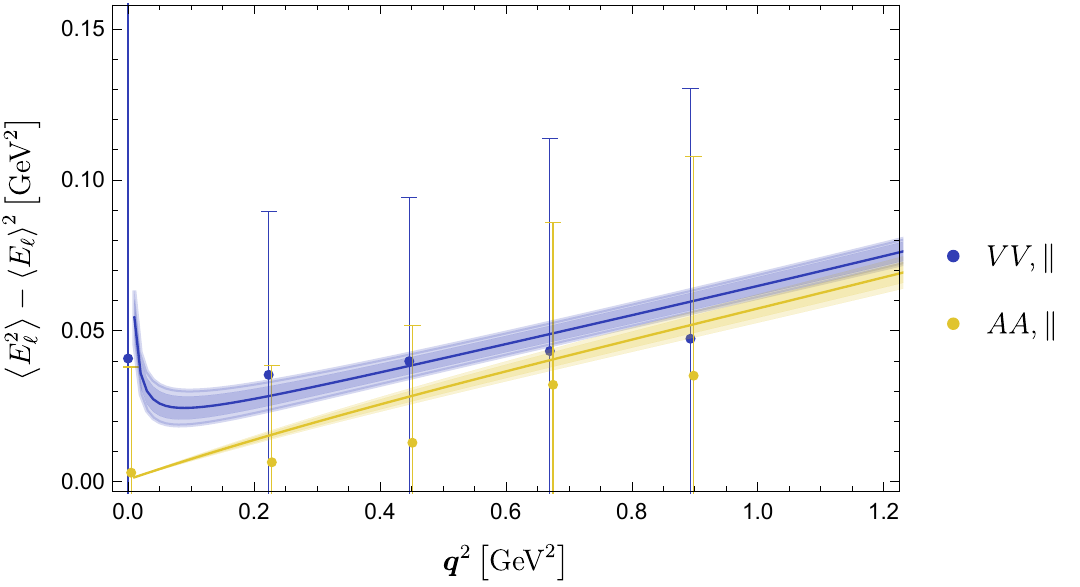}
    \end{subfigure}
    \begin{subfigure}{0.495\textwidth}
      \includegraphics[width=\textwidth]{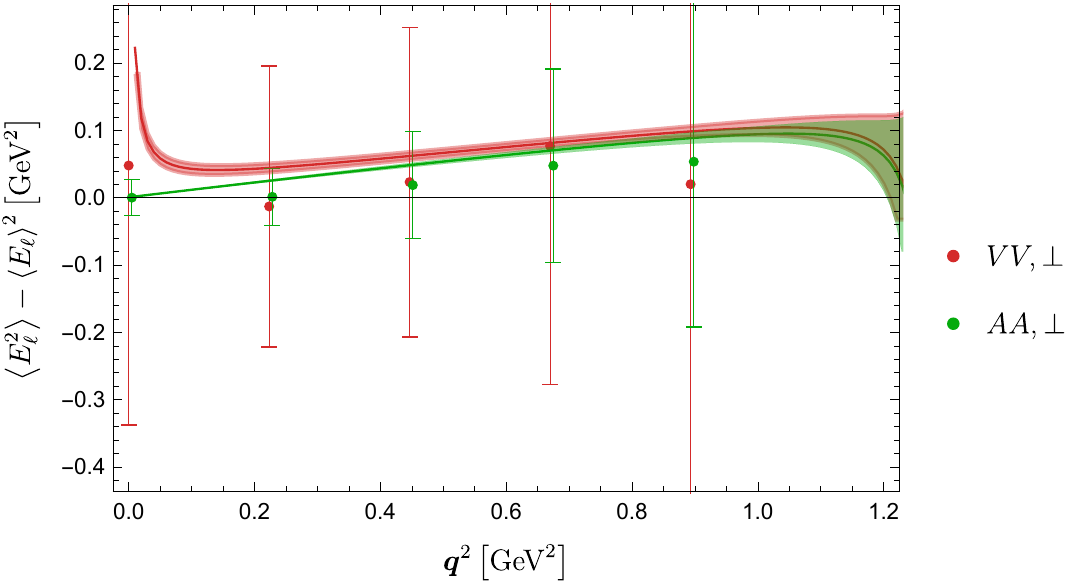}
    \end{subfigure}
      \caption{\label{fig:VariancePlot} This plot shows the variance $\sigma^2_{AA,\parallel} \left(\bm{q}^2 \right) $. The solid lines correspond to the OPE predictions including corrections up to $\mathcal{O}\left(1/m_b^3,\alpha_s\right)$. The dark bands around the OPE lines show the theoretical uncertainties arising due to missing higher order corrections and the light shaded bands additionally include the uncertainties originating from the fit used to determine the charm quark mass and the heavy quark matrix elements. The dots correspond to the lattice data.}
    \end{figure}

    In \cref{fig:VariancePlot} a comparison of the variance of the charged lepton energy can be found. Also in this case there is a good agreement between lattice data and OPE predictions. The current lattice uncertainties for this observable are rather large however.

\section{Conclusions}
  \label{sec:Conclusion}
  We have expanded the pilot study performed in \cite{Gambino:2020crt} by including corrections of $\mathcal{O}\left(1/m_b^3, \alpha_s \right) $, giving an estimate of the OPE uncertainties and considering lepton energy moments as additional observables. In the scenario of an unphysically light $b$ quark we find a remarkable agreement between lattice data and OPE predictions. 

  In order to obtain more information about the non-perturbative power corrections it might be useful to consider hadronic observables such as the moments of the hadronic final state invariant mass as well. We are currently preparing an extension of this work including also these moments and lattice data obtained by the ETMC collaboration.

\acknowledgments
  We thank the members of the JLQCD collaboration for helpful discussions and providing the computational framework and lattice data. Numerical calculations are performed on SX-Aurora TSUBASA at High Energy Accelerator Research Organization (KEK) under its Particle, Nuclear and Astro Physics Simulation Program, as well as on Oakforest-PACS supercomputer operated by Joint Center for Advanced High Performance Computing (JCAHPC). This work is supported in part by JSPS KAKENHI Grant Number 18H03710 and by the Post-K and Fugaku supercomputer project through the Joint Institute for Computational Fundamental Science (JICFuS). PG and SM are supported in part by the Italian Ministry of Research (MIUR) under grant PRIN 20172LNEEZ. This project has received funding from the Swiss National Science Foundation (SNF) under contract 200020\_204428.

\end{document}